\documentclass{article}

\usepackage{arxiv}

\usepackage[utf8]{inputenc} 
\usepackage[T1]{fontenc}    
\usepackage{hyperref}       
\usepackage{url}            
\usepackage{booktabs}       
\usepackage{amsfonts}       
\usepackage{nicefrac}       
\usepackage{microtype}      
\usepackage{lipsum}		
\usepackage{graphicx,url}
\usepackage[sort&compress]{natbib}
\usepackage{doi}
\usepackage{multirow}
\usepackage{units}
\usepackage{color, colortbl}
\usepackage{caption}
\usepackage[list=true]{subcaption}
\usepackage{enumitem}
\usepackage{comment}
\usepackage{float}
\usepackage{fdsymbol}
\usepackage{color, colortbl}

\title{Song Emotion Recognition: a Performance Comparison Between Audio Features and Artificial Neural Networks}

\author{ \href{https://orcid.org/0000-0002-8118-4213}{\includegraphics[scale=0.06]{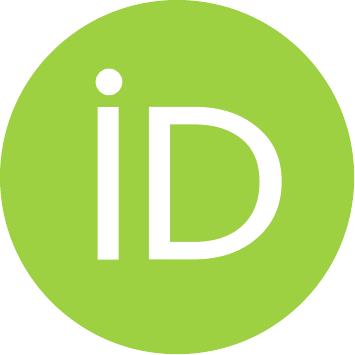}\hspace{1mm}Karen Gissell Rosero Jácome} \\
    School of Electrical and Computer Engineering\\
	University of Campinas, Brazil \\
	\texttt{k264373@dac.unicamp.br} \\
	\And
	\href{https://orcid.org/0000-0002-3989-7105}{\includegraphics[scale=0.06]{orcid.pdf}\hspace{1mm}Arthur Nicholas dos Santos} \\
	School of Electrical and Computer Engineering\\
	University of Campinas, Brazil \\
	\texttt{a264372@dac.unicamp.br} \\
	\And
	\href{https://orcid.org/0000-0002-5554-1483}{\includegraphics[scale=0.06]{orcid.pdf}\hspace{1mm}Pedro Benevenuto Valadares} \\
	School of Electrical and Computer Engineering\\
	University of Campinas, Brazil \\
	\texttt{p204483@dac.unicamp.br} \\
	\And
	\href{https://orcid.org/0000-0002-2246-4450}{\includegraphics[scale=0.06]{orcid.pdf}\hspace{1mm}Bruno Sanches Masiero} \\
	School of Electrical and Computer Engineering\\
	University of Campinas, Brazil \\
	\texttt{masiero@unicamp.br} \\
}

\renewcommand{\shorttitle}{SER: a Performance Comparison Between Audio Features and ANNs}

\hypersetup{
pdftitle={A template for the arxiv style},
pdfsubject={q-bio.NC, q-bio.QM},
pdfauthor={David S.~Hippocampus, Elias D.~Striatum},
pdfkeywords={First keyword, Second keyword, More},
}

\begin{document}
\maketitle

\begin{abstract}
	When songs are composed or performed, there is often an intent by the singer/songwriter of expressing feelings or emotions through it. For humans, matching the emotiveness in a musical composition or performance with the subjective perception of an audience can be quite challenging. Fortunately, the machine learning approach for this problem is simpler. Usually, it takes a data-set, from which audio features are extracted to present this information to a data-driven model, that will, in turn, train to predict what is the probability that a given song matches a target emotion. In this paper\footnote[1]{This work was supported by the São Paulo Research Foundation (FAPESP), grants \mbox{$\#$2017/08120-6}, $\#$2019/22795-1, $\#$2019/22945-3 and \mbox{$\#$2020/10195-7}. The opinions, hypothesis and conclusions or recommendations expressed in this material are the authors' responsibilities, and not necessarily reflect FAPESP's views.}, we studied the most common features and models used in recent publications to tackle this problem, revealing which ones are best suited for recognizing emotion in \emph{a cappella} songs.
\end{abstract}

\keywords{Song Emotion Recognition \and Chromagram \and Convolutional Neural Networks \and Recurrent Neural Networks \and Long Short-Term Memory}

\section{Introduction}
Music is art, which is a form of expression. By the time it reaches an audience, a spectrum of emotional reactions can be provoked, in account that Music Emotion Recognition (MER) is a process that is highly intertwined with people’s life experiences and cognitive capacities. In contrast, as a sub-field of Music Information Retrieval (MIR), MER deals with classification of music according to affective computing~\cite{kim2010music}. 

Currently, the importance of MER can be justified by the dependency of search and recommendation engines on metadata which, in simple terms, is just data about data. For example, when a person uses a smartphone to take a picture of a cat, the data is the picture itself. However, a series of other information about that data is also recorded, e.g., the time, date and geographical coordinates where it was shot etc. All of which are metadata that can be used to tag the data itself, to either retrieve it in the future or find similar content. Likewise, song metadata, e.g., (genre, composer, artist, album, year of release etc.) are commonly used by streaming services to help users to find what they may be prone to like, or even recommend songs based on their listening history. However, the mood of a song is also an interesting metadata, that could be used to relate a certain song to similar content.

In this paper, we studied articles with \textit{state-of-the-art} results published on MER, for both song and instrumental music. Our findings reveal that, usually, timbral features, e.g., Mel spectrogram, Mel-Frequency \textit{Cepstral} Coefficients (MFCC) etc., are employed as front-ends, whilst regarding Artificial Neural Network (ANN) models, the Multi-Layer Perceptron (MLP) and Convolutional Neural Network (CNN) architectures are most commonly employed as back-end. However, when comparing the performance of different audio features and ANN models, our experiments showed that the chromagram, which is a harmonic feature, combined with either one-dimensional (1-D) or two-dimensional (2-D) CNN architectures yields even better results.

The remainder of this paper is organized as follows: Section \ref{sec2} details our findings on what are the most commonly used audio features to represent music samples to the most commonly used machine learning (ML) model architectures. Section \ref{sec3} describes our experiment, being synthesized different ANN models based on the information retrieved from the previous section. Section \ref{sec4} shows our results, comparing it to previous \textit{state-of-the-art} works that used a same data-set as we did. Finally, Section \ref{sec5} presents some pertinent considerations to conclude this study.

\section{Features and Models}\label{sec2}

According to~\cite{1}, musical dimensions can be related to emotions by a set of high-level features, namely: melody, harmony, rhythm, dynamics, tone color (timbre), expressivity, texture, form and vocals. On the other hand, computational features are considered low-level, because they only provide primitive descriptions by which individual high-level ones may be identified. 

\subsection{Audio features}

By reviewing the works of \cite{2,3,4,5,6,7,8,9,10,11}, we found $47$ different low-level computational features being used separately or concatenated to better represent training data-sets, depending on different ML architectures used. All these features are available \textit{off-the-shelf} on Python libraries and MATLAB toolboxes, and 6 of them were found to be used on $76.6\%$ of the publications reviewed, which are:

\begin{itemize}[leftmargin=*]
    \item \textit{Chromagram}: relates to harmony, i.e., the sound produced by the combination of various pitches and indicates energy distribution along a 12-dimensional vector (one for each semitone in the super-just harmonic scale, i.e., from A to G$\#$.);
    
    \item \textit{Mel spectrogram}: relates to tone color, i.e., timbre, and decomposes an audio signal into a series of frequency channels inspired by the human cochlea, enabling to study the signal’s frequency distribution into so-called critical bands;
    
    \item \textit{Mel-Frequency Cepstral Coefficients (MFCC)}: also relates to tone color and measures spectral shape. Can be derived from a log magnitude Mel spectrogram based on the Discrete Cosine Transform (DCT). Typically, only the first 8 to 13 MFCCs are used for voiced signals;
    
    \item \textit{Spectral centroid}: also relates to tone color and represents the mean of the magnitude spectrum of the Short-Time Fourier Transform (STFT);

    \item \textit{Spectral roll-off}: also relates to tone color and indicates the frequency below which approximately $85\%$ of the magnitude spectrum distribution is concentrated;
    
    \item \textit{Zero-Crossing Rate (ZCR)}: also relates to tone color and represents the number of times a waveform changes sign in a window, indicating change of frequency and noisiness.
    
\end{itemize}
As for the other $41$ audio features (which were used on the other $23.4\%$ of the publications reviewed), $13$ of them were related to rhythm, $10$ were kindred to tone color, $6$ were cognated to harmony, $5$ were related to melody, 5 were related to dynamics, only $1$ was connected to texture, 1 to musical form and 1 to vocals.

\subsection{ML models}

According to~\cite{2}, what dictates which and how many features can be used as front-end for an ML model is the architecture of the model itself. By reviewing the works of \cite{2,3,4,5,6,7,8,9,10,11}, we found $12$ different architectures being used, separately or combined. $3$ of these were found to be used on $17\%$ of the publications reviewed, namely: Support Vector Machine (SVM), Multi-Layer Perceptron (MLP), and Convolutional Neural Network (CNN); and another $2$ on $10\%$ of the publications reviewed, being: Recurrent Neural Network (RNN) with Long Short-Term Memory (LSTM) blocks, and random forest. Since the initial purpose of our work was to compare the performance of ANN models, we opted to leave out SVM and random forest, which are ML models that are not based on ANNs, thus focusing on the remaining 3 models:

\begin{itemize}[leftmargin=*]
    \item \textit{MLP}: is a type of ANN that models the relationship between a set of training data and a group of known targets. Its architecture is based on a simplified understanding of how the human brain responds to stimuli from sensory organs and is best suited to problems where the relationship between input and output data is well understood, yet the process that relates both is extremely complex;
    
    \item \textit{CNN}: is a type of ANN based on convolutional operations that can extract high-level features from 1-D or 2-D low-level features. It can deeply extract underlying features contained in each frame, while retaining time-series features in the same direction. In classification problems, an MLP layer is usually employed at the end of a CNN architecture, to output its predictions;
    
    \item \textit{RNN-LSTM}: is a type of ANN which commonly relies on sequential data, i.e., time-series vectors. In the special case of an LSTM block, it can learn dependencies in the time-dimension of Time-Frequency (T-F) features, incorporating local context in ANN predictions. Hence, when trained on top of CNNs, for instance, the input data is no longer an individual \textit{``image''}, e.g., a spectrogram, but rather a sequence, more like in a \textit{``movie''}. Moreover, when using a bidirectional LSTM (Bi-LSTM) block, it can also handle the forward and backward flow of information in ANNs.
    
\end{itemize}
As for the other 7 architectures, K-Nearest Neighbors (K-NN) was found to be used on $7\%$ of the publications reviewed, RNN with Gated Recurrent Unit (GRU) blocks, decision tree (CART and C4.5) and State Vector Regressor (SRV) were found to be used on $4\%$ of the publications reviewed (each) and logistic regression was set to be used on $3\%$ of the publications reviewed.

\section{Experimental setup}\label{sec3}

To experiment with the aforementioned features and models, a portion of the Ryerson Audio-Visual Database of Emotional Speech and Song (\href{https://www.kaggle.com/datasets/uwrfkaggler/ravdess-emotional-speech-audio}{\texttt{RAVDESS}}) data-set was chosen, comprising $1,\!012$ audio-only files of song recordings, performed by $24$ actors, singing $2$ lexically matched statements in a neutral North American accent. Song emotions include neutral, calm, happy, sad, angry, and fearful expressions~\cite{12}. To extract audio features from this data-set, the Python package for music and audio analysis~\href{https://zenodo.org/record/3955228#.YrIMLHbMJPY}{\texttt{Librosa}} (version $0.8.0$) was used, and all ANN models were synthesized using~\href{https://www.tensorflow.org/}{\texttt{TensorFlow}} (version $2.2.0$), compiled using ADAM optimization, \textit{categorical cross-entropy} as loss function, and trained using a NVIDIA TITAN V Graphics Processing Unit (GPU).

\subsection{MLP model}\label{mlp}

\begin{figure}[h!]
\centering
\includegraphics[width=.98\textwidth]{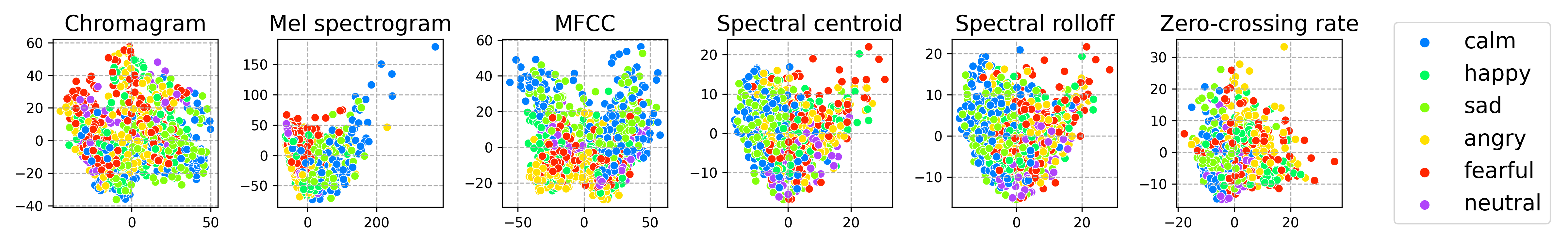}
\caption{PCA plots for all 6 audio features analyzed in this paper.}
\label{fig:pca}
\end{figure}

Since an MLP model can have an input layer with as many neurons as necessary, all input features could be concatenated and flattened into a $1$-D input vector. After extracting all $6$ audio features from the data-set, the Principal Component Analysis (PCA) technique was used to visualize the minimum number of variables that keeps the maximum amount of information about how each feature data is distributed, as illustrated in Figure~\ref{fig:pca}. 

Since Mel spectrogram showed the worst clustering and the highest dimensionality compared to the other features, we chose to train an MLP model using the concatenation of the other $5$ features, i.e., chromagram, MFCC, spectral centroid, spectral roll-off and ZCR. Hence, our MLP model consisted of an input layer with $11,\!394$ neurons, followed by $2$ hidden layers with $1,\!024$ and $128$ neurons each, totaling over $141$M free parameters. Rectified Linear Unit (ReLU) was used as an activation function for all layers, except for the output one, where \textit{softmax} was used. For regularization, we used \textit{dropout}, which randomly ignores a percentage of neurons during training.

\subsection{2-D CNN model}

For our 2-D CNN model, we selected the $2$-D feature that showed the clearest clustering in the PCA visualization in Figure~\ref{fig:pca}, i.e., the chromagram. Thus, the input for this architecture had a shape of $(12, 422, 1)$, followed by three convolutional blocks, with $24$, $48$, and $48$ filters respectively, \textit{kernel} sizes of $(5,5), (2,2), (3,3)$, convolutional layers' stride sizes of $(1,1)$, and \textit{max pooling} layers in the first two blocks, with $(2,4), (1,3)$ \textit{pooling} sizes and stride sizes of the same dimensions respectively. After the last convolutional block, the data was flattened into a $1$-D vector with $1,\!536$ elements, to be fed into $2$ Fully Connected (FC) layers with $64$ and $6$ neurons respectively, totaling almost 125k free parameters. For regularization, a \textit{dropout} of $0.5$ was used before each FC layer. ReLU was used as the activation function for all layers, except for the output one, where \textit{softmax} was used, to obtain the probability associated with each emotion. 

\subsection{1-D CNN model}\label{1dcnn}

To assess the performance of $1$-D CNN architectures, we reshaped the chromagram feature to match a model with an input length of $5064$ samples. Following this, $3$ convolutional blocks were added, with $16$, $32$ and $32$ filters respectively and kernel size of $4$. The \textit{max pooling} operation was implemented for all blocks with a \textit{pooling} size of $3$ and a stride of $1$. The data outputted by the last convolutional block was flattened before being passed on to $3$ FC layers with $1024$, $128$ and $6$ neurons, respectively, totaling over 6M free parameters. Regularization and activation functions were maintained as in the 2-D CNN model.

\subsection{Convolutional recurrent neural network (CRNN) model}

Finally, we also experimented with the addition of a Bi-LSTM block after the convolutional layers of the model described in Section \ref{1dcnn}, aiming to learn with the long short-term temporal dependencies of audio signals. The convolutional blocks have the same parameters as described before, except for the number of filters, which are $16$ in every block. The bidirectional wrapper takes an LSTM layer as argument with $100$ memory units, using ReLU as activation function and a \textit{dropout} of $0.5$ as regularization. Ultimately, the Bi-LSTM layer output is flattened and passed on to $3$ FC layers with the same characteristics detailed in Section~\ref{1dcnn}, totaling over 19M free parameters.

\subsection{Data augmentation}

In this work, we also explored suitable Data Augmentation (DA) techniques, i.e., creating slightly modified new data derived from the original one. Since ANNs consider these new data as genuine, they can benefit from it, learning new parameters to achieve even better performances without over-fitting. For this reason, we used the \href{https://github.com/iver56/audiomentations}{\texttt{Audiomentations}} library to add Gaussian noise to the original audio samples, also shift the song's pitches by $1$ or $2$ octaves. Models trained \underline{without} DA were trained for 100 epochs, while models trained \underline{with} DA were trained for 200 epochs.

\section{Results and discussion}\label{sec4}

To train our models, the data-set was split into 612 samples for training, 200 for validation and 200 for test.  In a previous work (\cite{dos2021song}), it was already evinced that 2-D CNN models can surpass the overall test accuracy of MLP models. Therefore, our MLP model was trained only without data augmentation, first using the concatenation of $5$ features described in Section \ref{mlp} and then in a feature ablation manner, as detailed in Table~\ref{results}. Next, our 2-D CNN model was trained with and without DA to compare its performance under these different conditions. Since the use of DA improved the system's metrics as evinced in Table~\ref{results}, we continued using DA for the training of our 1-D CNN and CRNN architectures, always saving the epoch that achieved the best validation accuracy. The results show that the 1-D and 2-D CNN models achieved the best performances compared to all other models, while taking less time to train, compared with our CRNN model. 

Moreover, we also resorted to the works of \cite{13,14,15} to compare our results with other works that used the same data-set, but not necessarily the same features and model architectures.

According to \cite{priore2014subversive}, harmony can be effectively used in songwriting to  encode hidden meanings, e.g., to imprint emotiveness regardless of rhythm, lyrics etc., which is corroborated by our findings in terms of accuracy. However, according to \cite{14, 15}, we were expecting that adding a Bi-LSTM cell would improve the accuracy of our 1-D CNN model, since a same musical scale sang in different orders can imprint different moods. In spite of that, our CRNN model was significantly outperformed by our 1-D and 2-D CNN models, with DA. 
\definecolor{Gray}{gray}{0.9}
\definecolor{LightCyan}{rgb}{0.9961,0.8471,0.6941}
\begin{table}[H]
\centering
\caption{Comparative results for different audio features and ANN models, using the song portion of the RAVDESS data-set.}
\footnotesize
\begin{tabular}{| c | c | c | c | c | c | c |}
\hline
\rowcolor{LightCyan} \textbf{Model} & \textbf{Features} & \textbf{Val. acc.} & \textbf{Val. loss} & \textbf{Test acc.} & \textbf{Test loss} & \textbf{Tr. time} \\ \hline 
\multirow{6}{*}{MLP$^{\varheartsuit}$} & Chromagram & 0.78$\pm$0.07 & 1.5$\pm$0.01 & 0.77$\pm$0.03 & 1.54$\pm$0.08 & \unit{18.24}{s} \\ \cline{2-7} 
 &  MFCC & 0.70$\pm$0.03 & 3.71$\pm$0.11 & 0.70$\pm$0.01 & 3.04$\pm$0.97 &  \unit{28.06}{s} \\ 
\cline{2-7} & Spec. centroid & 0.44$\pm$0.02 & 3.15$\pm$1.11 & 0.44$\pm$0.02 & 3.17$\pm$1.27 & \unit{3.13}{s} \\ 
\cline{2-7} & Spec. roll-off & 0.45$\pm$0.01 & 2.91$\pm$1.01 & 0.45$\pm$0.02 & 2.91$\pm$0.95 & \unit{3.03}{s} \\ 
\cline{2-7} & ZCR & 0.43$\pm$0.03 & 2.99$\pm$0.25 &0.38$\pm$0.07 & 3.03$\pm$0.48 & \unit{3.18}{s} \\ \cline{2-7} 
&  $5$ feats. concat. & 0.65$\pm$0.06 & 2.5$\pm$0.76 & 0.65$\pm$0.09             & 2.66$\pm$1.74 & \unit{5.14}{min} \\ 
\cline{1-7} \multirow{4}{*}{2-D CNN$^{\varheartsuit}$} & Mel spec. & 0.78$\pm$0.01 & 0.87$\pm$0.17 & 0.78$\pm$0.07 & 0.79$\pm$0.11                                   &           \unit{16.74}{s} \\ 
\cline{2-7} &  MFCC & 0.70$\pm$0.01 & 0.81$\pm$0.04 & 0.63$\pm$0.02 & 0.86$\pm$0.01 & \unit{7.74}{s} \\ \cline{2-7} & Chromagram & 0.84$\pm$0.01 & 0.49$\pm$0.02 & 0.80$\pm$0.01 & 0.62$\pm$0.02      & \unit{8.42}{s} \\ 
\cline{2-7} & \begin{tabular}[c]{@{}c@{}}Chromagram / \\ DA\end{tabular} & 0.80$\pm$0.01 & 0.62$\pm$0.05 & \textbf{0.84}$\pm$0.02 & 0.57$\pm$0.08 & \unit{17.66}{s} \\ 
\cline{1-7} 1-D CNN$^{\varheartsuit}$ & \begin{tabular}[c]{@{}c@{}}Chromagram / \\ DA\end{tabular} & 0.87$\pm$0.02 & 0.42$\pm$0.07       & 0.83$\pm$0.01 & \textbf{0.42}$\pm$0.05 & \unit{44.82}{s} \\ 
\cline{1-7} \begin{tabular}[c]{@{}c@{}}1-D CNN \\ + BiLSTM$^{\varheartsuit}$\end{tabular}      &  \begin{tabular}[c]{@{}c@{}}Chromagram / \\ DA\end{tabular} & 0.81$\pm$0.03 & 0.81$\pm$0.45 & 0.75$\pm$0.02 & 0.89$\pm$0.44            & \unit{11.21}{min} \\ \hline
\begin{tabular}[c]{@{}c@{}}Logistic \\ Regression$^{\vardiamondsuit}$\end{tabular}      &\begin{tabular}[c]{@{}c@{}}MATLAB Audio \\ Analysis Library\end{tabular}      &          -             &               -    & 0.48                &          -                                  &             -           \\ \cline{1-7}
RNN (LSTM)$^{\spadesuit}$ &  Librosa HSF      &            -           &        -           & 0.82                &                     -                       &                    -    \\ \hline
\begin{tabular}[c]{@{}c@{}}1-D CNN \\ + BiLSTM$^{\clubsuit}$\end{tabular}        & \begin{tabular}[c]{@{}c@{}}MFCC / \\ DA\end{tabular}             &            -           &           -        & 0.73                &                 -                      &      -                  \\ \hline
\end{tabular}
\label{results}
\scriptsize{\\Results associated with: $^{\varheartsuit}$\textbf{ours}, $^{\vardiamondsuit}$\cite{13}, $^{\spadesuit}$\cite{15} and $^{\clubsuit}$\cite{14}. All our models were trained using k-fold cross-validation, from which the mean and standard deviation values shown were computed.}
\end{table}
\begin{figure}[H]
\centering
\includegraphics[width=.88\textwidth]{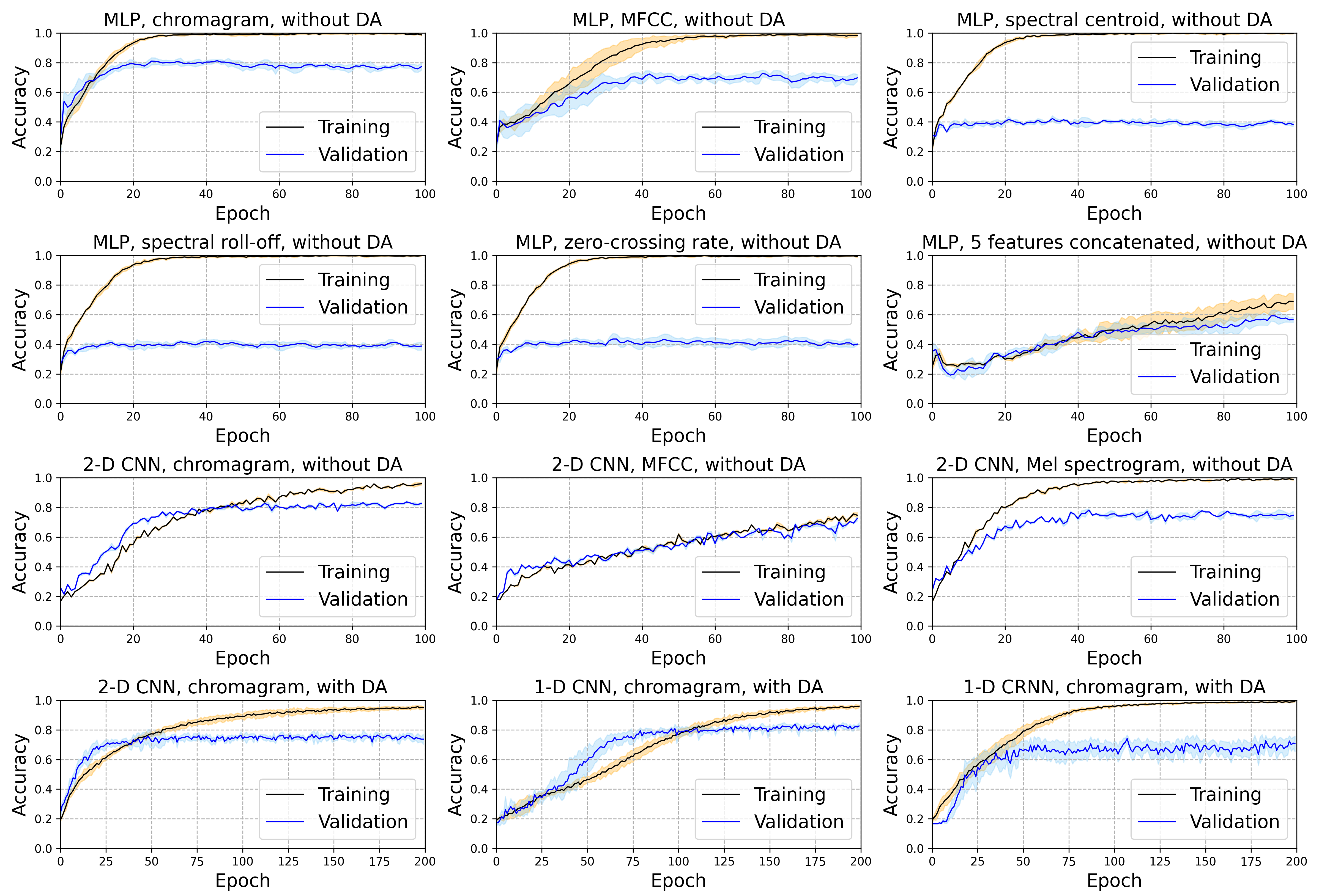}
\caption{Learning curves for all 12 combinations of features and models tested. Shadows enveloping the curves represent standard deviations.}
\label{fig:lc}
\end{figure}

In terms of training time, our CRNN model was also outperformed by our 1-D and 2-D CNN models, and while our MLP models took even less time to train, they also produced lower accuracy in comparison. Looking into the learning curves of all 12 combinations of audio features and ANN models tested as illustrated in Figure \ref{fig:lc}, it's observable that, regarding the MLP models (without DA), the ones with chromagram and MFCC as front-end start to diverge at around epoch 20 and start to over-fit at around epoch 60. On the other hand, the ones with spectral centroid, spectral roll-off and ZCR as front-end start to diverge early in their training, and over-fit at around epoch 20, which is consistent with their poor performances. 

Using $5$ features concatenated as front-end, the MLP model achieves its best accuracy, which can be justified by its late divergence. Regarding the 2-D CNN models, the ones with chromagram and Mel spectrogram as front-end start to diverge early in their training, yet stabilizing at high values of validation accuracy earlier as well. With DA, the 2-D CNN model stabilizes even earlier than others, which justifies its good performance. Finally, regarding the 1-D CNN models, it's noticeable that, without the Bi-LSTM layer, the model stabilizes at higher values of validation accuracy, without ever fully over-fitting, contrary to its CRNN counterpart.

\section{Conclusions}\label{sec5}

Although the most popular computational features for MER used in recent publications relate to tone color, in our experiments the chromagram, which relates to harmony, was found to be best suited for song emotion recognition. Also, our 1-D and 2-D CNN models performed better than both our MLP and CRNN models, despite earlier works showing that RNN layers should improve the accuracy of 1-D CNN models. Moreover, our best result ($0.84\pm0.02$ test accuracy) was obtained using data augmentation, evincing that ANNs do benefit from this technique. Ultimately, regarding \underline{song} emotion recognition, our results are \textit{state-of-the-art}, compared with recent publications, but it's important to mention that the data-set used in our experiments is far from being a general representation of the human population. Since it's samples lack accent, language, ethnical and gender diversities, disability inclusion etc., when put to test with samples different from its own \textit{corpus}, it will hardly produce exciting results, such as in Table \ref{results}.

\bibliographystyle{unsrt}







\end{document}